\documentclass[12pt]{iopart}
\usepackage{graphicx}

\graphicspath{{../fig/}}

\usepackage{color}

\newcommand{\removed}[1]{}

\usepackage{gensymb}

\begin{document}

\title{Dendritic flux avalanches in a superconducting MgB$_2$ tape}

\author{T Qureishy$^1$, C Laliena$^2$, E Mart\'{i}nez$^2$, A J Qviller$^3$, J I Vestg{\aa}rden$^1$ $^4$, T H Johansen$^1$ $^5$, R Navarro$^2$ and P Mikheenko$^1$}
\address{$^1$ Department of Physics, University of Oslo, P. O. Box 1048 Blindern, 0316 Oslo, Norway}
\address{$^2$ Instituto de Ciencia de Materiales de Arag\'{o}n, (CSIC - Universidad de Zaragoza), C/ Mar\'{i}a de Luna 3, 5018 Zaragoza, Spain}
\address{$^3$ nSolution AS, Maries gate 6, 0368 Oslo, Norway}
\address{$^4$ Norwegian Defence Research Establishment (FFI), Kjeller, Norway}
\address{$^5$ Institute for Superconducting and Electronic Materials, University of Wollongong, Northfields Avenue, Wollongong, NSW 2522, Australia}

\ead{thomashq@fys.uio.no}

\begin{abstract}
	MgB$_2$ tapes with high critical current have a significant technological potential, but can experience operational breakdown due to thermomagnetic instability. Using magneto-optical imaging the spatial structure of the thermomagnetic avalanches has been resolved, and the reproducibility and thresholds for their appearance have been determined. By combining magneto-optical imaging with magnetic moment measurements, it is found that avalanches appear in a range between 1.7 mT and 2.5 T. Avalanches appearing at low fields are small intrusions at the tape's edge and non-detectable in measurements of magnetic moment. Larger avalanches have dendritic structures. 
\end{abstract}

\noindent{\it Keywords}: superconductivity, MgB$_2$ tapes, magneto-optical imaging, dendritic avalanches

\maketitle

\section{Introduction}

Soon after the discovery of superconductivity below 39 K in magnesium diboride (MgB$_2$) \cite{Nagamatsu2001}, the huge interest of the scientific community led to an extensive physical characterization of this type-II superconductor \cite{0953-2048-14-11-201,0953-2048-22-4-043001}. Improved magnetic flux pinning and hence a higher critical current density, $J_{\rm c}$, was achieved with carbon substitution for boron by chemical doping \cite{Dou2002_APL81}. 

The powder-in-tube technology, extensively used in the fabrication of the first-generation high-temperature superconductors, was found to be useful for producing MgB$_2$ tapes with high $J_{\rm c}$ and significant technological potential \cite{0953-2048-14-4-304}. Iron is often used as metallic sheath material for MgB$_2$ wires and tapes because of its good mechanical properties and chemical compatibility with MgB$_2$ \cite{Jin2001,Suo2001}. 

Magnetic flux jumps are observed in most type-II superconductors when applying a magnetic field at low temperatures. These abrupt events are caused by a thermomagnetic instability \cite{RevModPhys.53.551,WIPF1991936}. Magnetic vortices move into the superconductor as an applied magnetic field or current is changed. When moving, they dissipate heat, and if this heat is not removed quickly, more vortices depin and move into the sample dissipating even more heat, and this self-amplifying process results in a thermomagnetic avalanche. Such events lead to jumps in magnetic moment of bulk MgB$_2$ \cite{Dou200179}, as well as in MgB$_2$ films \cite{0295-5075-59-4-599}, wires \cite{Wang2001149} and tapes \cite{0953-2048-17-3-L02}. 

The spatial structure of flux avalanches can be observed by magneto-optical imaging (MOI). Most commonly they have a dendritic structure, as found in thin films made of Nb \cite{PhysRevB.52.75}, NbN \cite{doi:10.1063/1.1992673}, MgB$_2$ \cite{0295-5075-59-4-599,0953-2048-14-9-319} and YBCO \cite{PhysRevLett.71.2646,0295-5075-64-4-517}, and also in foils of Nb \cite{WERTHEIMER19672509}. The detailed structure of dendritic avalanches are unpredictable. Yet, they have upper and lower thresholds for both increasing and decreasing magnetic fields \cite{Qviller2010897}. In films of MgB$_2$ cooled in zero magnetic field, dendritic avalanches have a threshold temperature of 10 K when applying an increasing magnetic field \cite{0953-2048-14-9-319}.

By combining MOI with magnetic moment measurements, we have investigated flux jumps in a 50-$\mu$m thick carbon-doped MgB$_2$ tape. We report observations of dendritic avalanches in the sample. Their structures, reproducibility, threshold magnetic fields and temperature-dependence are discussed.

\section{Methods}

Six MgB$_2$ tapes were synthesized and characterized by MOI and magnetic moment measurements. Dendritic avalanches were found in only one of them, and the present paper focuses on detailed results for this particular sample. MOI results from four other tapes are presented in \cite{Laliena2017}. An MgB$_2$ wire was synthesized as well, and characterized by magnetic moment measurements. 

The MgB$_2$ tape was manufactured by the powder-in-tube method using $in$ $situ$ reaction of ball-milled precursor powders, similarly to the process described in \cite{Laliena2017}, but with oleic acid added to the powders before milling to improve its critical current density. Powders of Mg and B were mixed in stoichiometric ratio (1:2) with oleic acid in a Retsch MM 200 vibratory ball mill for 30 minutes. The amount of oleic acid was 10 wt.\% of the Mg-B mixture. In order to obtain optimum superconducting properties, the precursor powder was subsequently heat treated at 400 \degree C in an argon atmosphere for an hour \cite{0953-2048-26-12-125017}. After that, the powder was milled in a Retsch PM 100 planetary ball mill with balls of tungsten carbide, rotated at a speed of 200 revolutions per minute for 1.5 hours. Every three minutes the milling was paused for one minute, and the rotation direction was changed. 

The resulting powder was inserted into an iron tube with an inner and outer diameter of 4.5 and 5 mm, respectively. The tube was then sealed at both ends. After that, it was drawn through the formers in consecutive steps reducing the diameter down to 1.1 mm and finally cold-rolled into a 2-mm wide and 0.4-mm thick tape. During mechanical deformation, intermediate annealing at 550 \degree C for an hour in argon atmosphere was performed to reduce the iron sheath's work-hardening. Finally, the tape was also sealed at both ends and annealed at 670 \degree C for five hours in vacuum to form the final product. 

After annealing, the tape was cut into several pieces for characterization with different techniques. For MOI analysis, which is the main focus in the present work, a 10 mm-long piece was used. For measurements, one side of the tape was polished to expose the superconducting core, resulting in the sample, approximately 50 $\mu$m thick. For comparison, a piece of the same MgB$_2$ wire, without the rolling process, was annealed with the same conditions as were used for the tape.  

For characterization, the tape was mounted in a helium flow cryostat with a magneto-optical indicator film placed on top of it. The indicator is a Faraday-active layer, which in an optical system with crossed polarizers allows one to visualize magnetic flux that penetrates the specimen \cite{0034-4885-65-5-202}. In our case it is a bismuth-substituted ferrite garnet film deposited on a gadolinium gallium garnet substrate with an aluminium mirror to reflect the light coming from the polarizer \cite{PhysRevB.64.174406,PhysRevB.66.064405}. The superconductor with the indicator on top was cooled in zero magnetic field to a temperature below its critical temperature, $T_{\rm c}$. Then a field was applied perpendicular to its surface and increased to $\mu_{\rm 0}H$ = 85 mT, while taking images every 0.85 mT. After that, the sample was heated above $T_{\rm c}$, and the process was repeated. 

4-5 mm long samples of the same MgB$_2$ wire and tape were characterized by vibrating sample magnetometers (VSM, Quantum Design PPMS-9T and PPMS-14T) and SQUID magnetometer (Quantum Design 5T). After removing the iron sheath by mechanical polishing, the diameter and thickness of the wire and tape were 0.9 mm and 0.29 mm, respectively. The isothermal magnetic hysteresis loops at $T$ = 5 K were measured. For the tape, a magnetic field was applied perpendicular to its flat surface and increased from $\mu_{\rm 0}H$ = 0 to 9 T. After that, it was decreased back to zero, and finally increased in the opposite direction to - 9 T. In the case of the wire, the field was applied perpendicular to its axis and increased from 0 to 14 T, decreased back to 0 and then increased in the opposite direction to - 14 T.

\section{Results and discussion}

\begin{figure}[t]
	\centering
	\includegraphics[width=8.0cm]{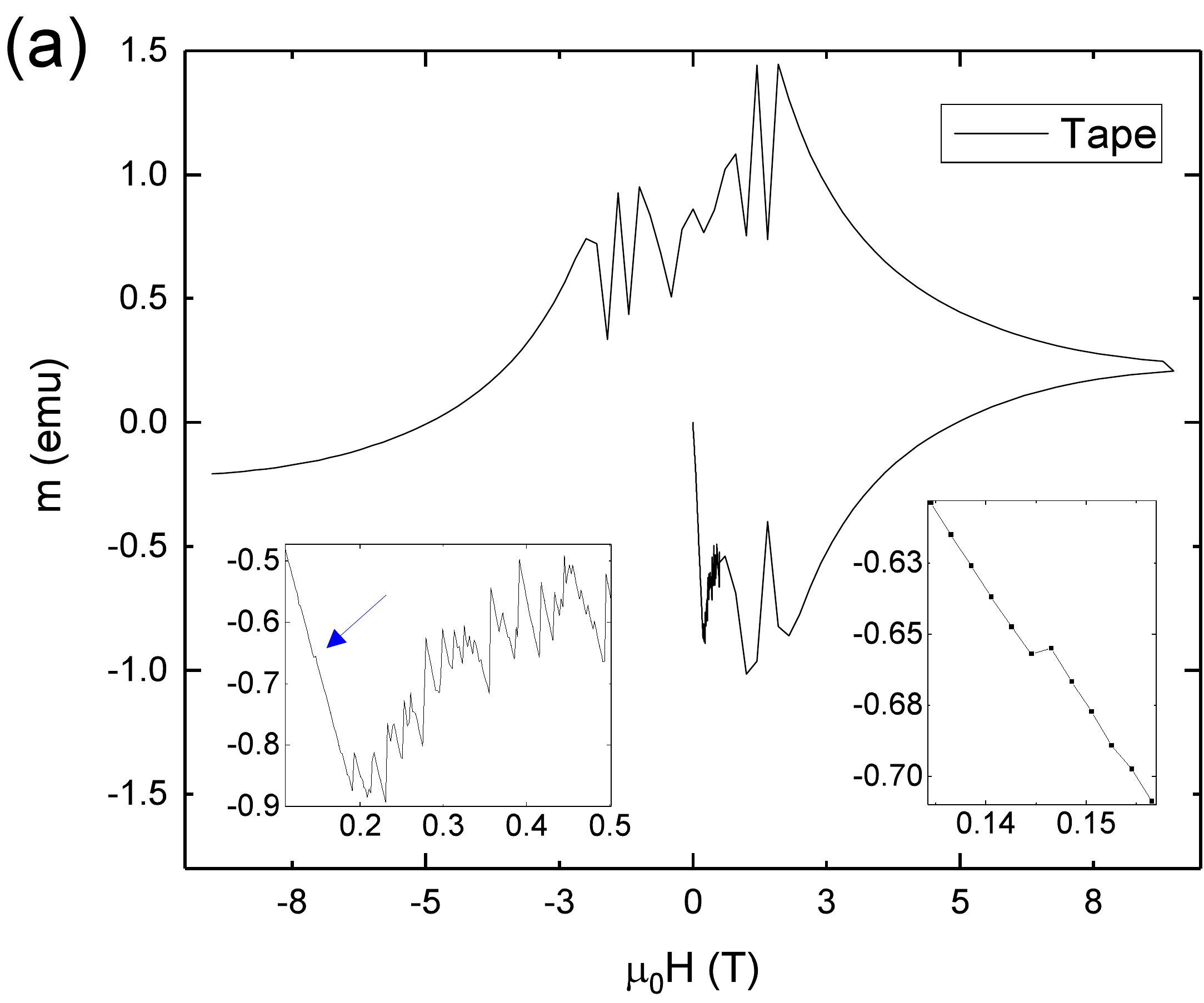} 
	\includegraphics[width=8.0cm]{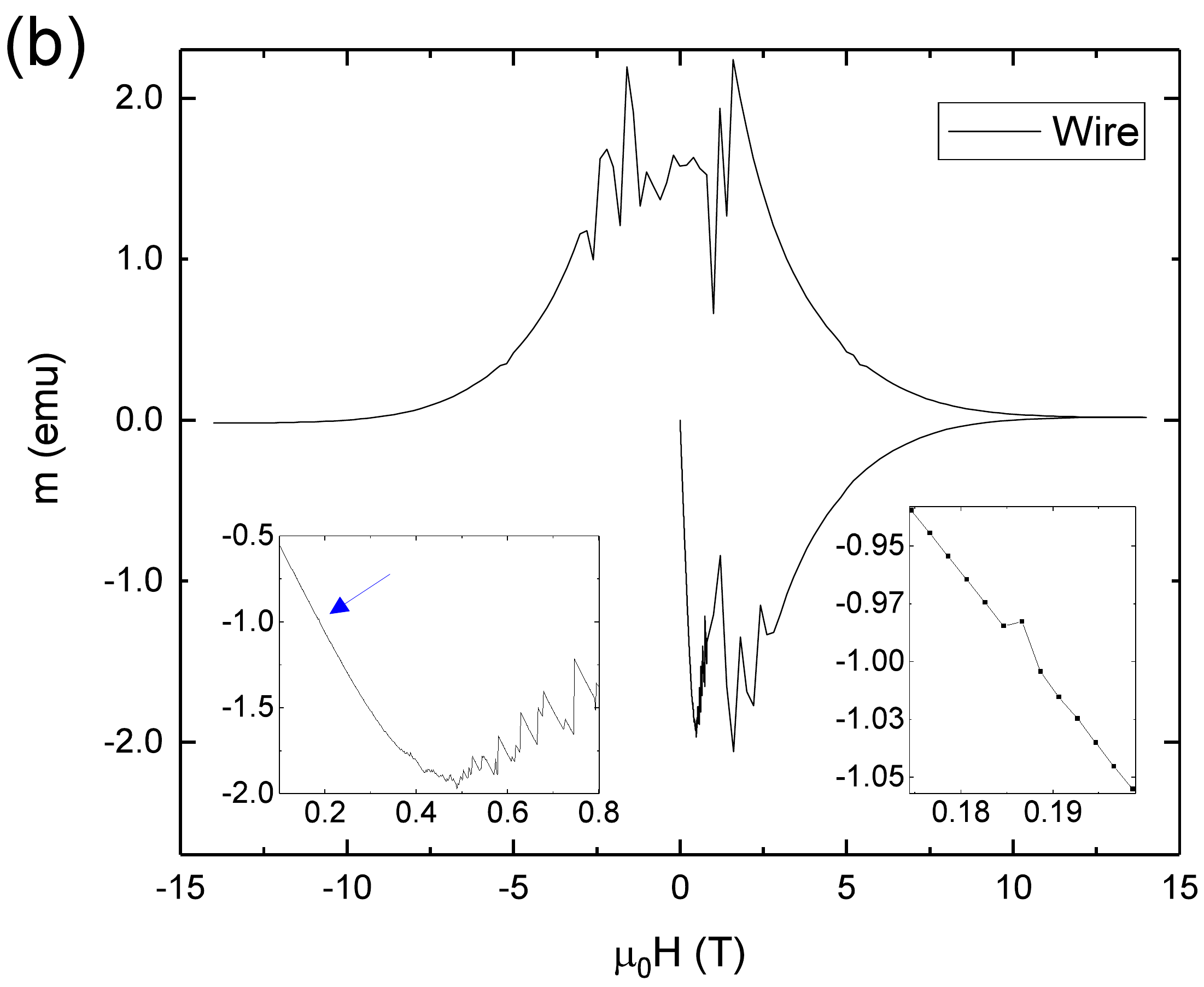} 
	\caption{
		\label{fig:mag}
		Magnetic moment as a function of applied magnetic field for (a) tape and (b) wire at a temperature of 5 K. The insets in the lower-left corner of both figures show several of the first flux jumps in the virgin branch. An arrow points at the position of the first flux jump, which is shown in the inset in the lower-right corner of both figures. The units on the axes in the insets are the same as those in the main figures. 
	}
	
\end{figure}

In figure~\ref{fig:mag}(a), magnetic moment $m$ of the tape is plotted as a function of applied magnetic field $\mu_{\rm 0}H$ at $T$ = 5 K. $m$ is smooth at high fields and contains flux jumps at low fields. Flux jumps are seen in the virgin curve and also in the reverse branches, after having applied the maximum field, in fields between + 2.5 and - 2.5 T. The inset in the lower-left corner shows several of the first flux jumps occurring in the virgin curve, including the first one at $\mu_{\rm 0}H$ = 0.145 T, marked by an arrow. The first flux jump is also shown in the inset in the lower-right corner.  

Figure~\ref{fig:mag}(b) shows $m$ as a function of $\mu_{\rm 0}H$ at 5 K for the MgB$_2$ wire. There are many flux jumps in the reverse branch in approximately the same field interval as in the tape, and also flux jumps up to about 3 T in the virgin curve of the wire. The inset to the left shows several of the first flux jumps, and an arrow points at the first one at $\mu_{\rm 0}H$ = 0.185 T. The inset to the right shows the first flux jump. 

The exact flux jump pattern in $m$($H$) measurements depends on the ramp rate $\mu$d$H$/d$t$, which is 13-30 mT/s for these measurements. Since both the wire and tape contain similar flux jump behaviour, it is likely that rolling the wire into the tape before the final annealing and polishing off the iron sheath is not the only cause of flux jumps in the tape. 

Since magnetic moment measurements give the average magnetic properties over the whole volume of the samples, it is desirable to visualize local magnetic fields by MOI to clarify the sample behaviour, even if the operating range of the indicator films is limited to low magnetic fields, in our case 85 mT, below the first detected jump in the magnetic moment measurements. 

\begin{figure}[t]
	\centering
	\includegraphics[width=8.0cm]{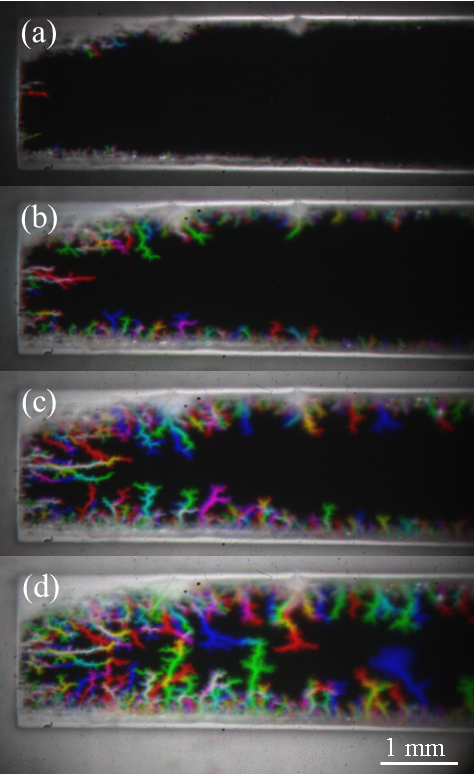} 
	\caption{
		\label{fig:RGBord}
		Colour coded MOI images obtained after zero-field cooling of MgB$_2$ tape to 3.7 K and application of increasing perpendicular magnetic field of (a) 18.7, (b) 42.5, (c) 62.1 and (d) 84.2 mT. The procedure was carried out three times, and each image is a superposition of three images, plotted in red, green and blue.  
	}
\end{figure}

Figure~\ref{fig:RGBord} shows colour-coded MOI images of a section of the tape covering two thirds of its whole length. The images were obtained at 3.7 K and different magnetic fields. Each image consists of a superposition of three images obtained under the same conditions (zero-field cooling to 3.7 K and applying the same magnetic field). They are presented in three different colours: red, green and blue (RGB). Grey colours in these images appear as a sum of these three colours. Brightness and contrast were optimized for each image individually. The images from the top to bottom correspond to increasing applied fields, namely to $\mu_{\rm 0}H$ = 18.7, 42.5, 62.1 and 84.2 mT, respectively. The inner region is dark, indicating zero flux, and the area outside of its borders is brighter, which shows that the tape expels magnetic field. The bright horizontal lines at the upper and lower edges of the tape are from the remnants of ferromagnetic sheath. Bright flux front propagating into the sample shows the advancement of magnetic flux. 

Magnetic flux penetrates gradually into the sample from the edges and forms a critical state-like region with a non-smooth flux front. The flux front is also non-homogeneous, especially in the upper-left corner, because of enhanced surface roughness in the tape there. In addition to gradual flux penetration, figure~\ref{fig:RGBord} clearly shows specific dendritic formations. These formations appear suddenly and have branch-like structures resembling lightning. The first dendrites are small, with very few branches. As the applied magnetic field increases, new and larger dendrites appear. At the highest applied fields, new ones have even more branches. Most of the dendrites have a non-reproducible appearance, as can be seen in the RGB images, where colours are not mixed. The parts of dendrites that are yellow, cyan or magenta were created by dendrites following the same path in two out of three measurements. This makes the pattern or parts of the pattern reproducible to some extent, although branches rarely follow the same path at the same applied fields. 

\begin{figure}[t]
	\centering
	\includegraphics[width=8.0cm]{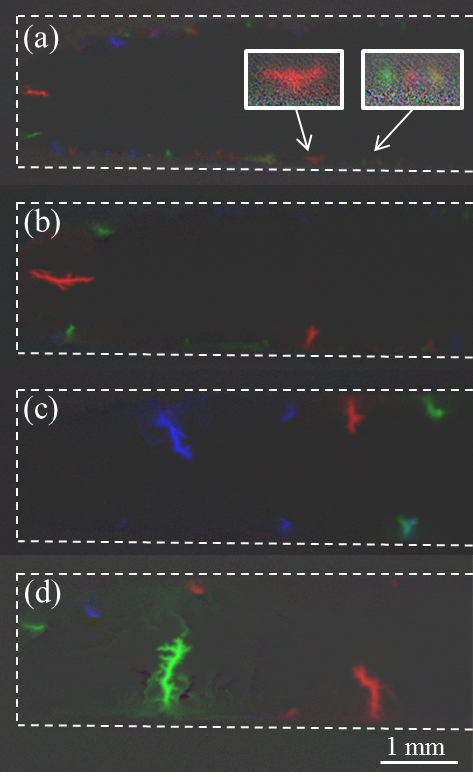} 
	\caption{
		\label{fig:RGBdiff}
		Differential RGB MOI images obtained after zero-field cooling the tape to 3.7 K and applying increasing perpendicular magnetic field. White dashed lines show the outline of the tape. The applied fields are the same as those in figure~\ref{fig:RGBord}, namely (a) 18.7, (b) 42.5, (c) 62.1 and (d) 84.2 mT, and the images are obtained by subtracting preceding images taken at slightly lower (by 0.85 mT) magnetic field. The two insets in (a) show magnified images with enhanced brightness of several avalanches appearing from the bottom edge of the sample. 
	}
\end{figure}

More information can be extracted from differential MOI images presented in figure~\ref{fig:RGBdiff}. In this figure, the temperature and fields are the same as in figure~\ref{fig:RGBord}, with the only difference that preceding images are subtracted from them, which were obtained at slightly lower (by 0.85 mT) magnetic field. White dashed lines in the figure show the outline of the tape and were added as a guide to the eye. The insets in figure~\ref{fig:RGBdiff}(a) show magnified images of two areas with enhanced brightness. As one can see, there are quite small dendritic avalanches at low applied magnetic fields, and several larger ones that appear at higher fields. The small dendrites at lower fields do not have as many branches as those appearing at higher fields. 

\begin{figure}[t]
	\centering
	\includegraphics[width=8.0cm]{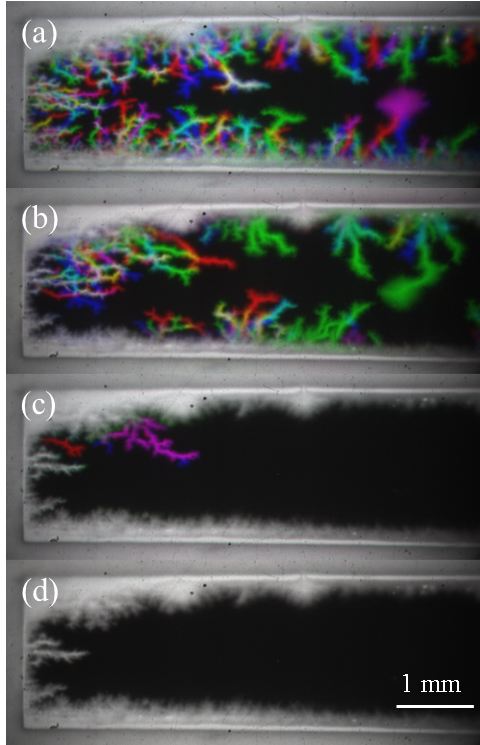}
	\caption{
		\label{fig:RGBtemp}
		RGB MOI images after zero-field cooling of the tape to (a) 4.0, (b) 6.0, (c) 8.0 and (d) 10.0 K and subsequently applying perpendicular magnetic field of 85 mT, three times at each temperature.  
	}
\end{figure}

Figure~\ref{fig:RGBtemp} shows images obtained at zero-field cooling to different temperatures of 4.0, 6.0, 8.0 and 10.0 K (from top to bottom), and applying magnetic field of 85 mT. Dendritic avalanches are seen in Figs.~\ref{fig:RGBtemp}(a)-(c), but not in figure~\ref{fig:RGBtemp}(d). The higher the temperature, the fewer dendritic avalanches are observed. The magenta colour of the dendrite in figure~\ref{fig:RGBtemp}(c), which is a mixture of red and blue, shows that this particular dendrite is almost reproducible in two experiments. On the left end of the tape in Figs.~\ref{fig:RGBtemp}(c) and~\ref{fig:RGBtemp}(d), there is a white dendrite-like formation. However, magnetic flux penetrates gradually there while increasing the field, which indicates that it is not a thermomagnetic instability, but merely flux penetration into a crack in the sample. This colour-less feature has a similar structure to the large dendritic avalanches, which indicates that cracks play an important role in the resulting pattern of dendritic avalanches. This is further supported by the similarity between the red dendrite in figure~\ref{fig:RGBdiff}(b) and the colour-less dendrite in Figs.~\ref{fig:RGBtemp}(c) and ~\ref{fig:RGBtemp}(d). The red avalanche may have propagated into the crack. This could explain the few cases of reproducibility of dendritic formations in the sample. 

\begin{figure}[h]
	\centering
	\includegraphics[width=8.0cm]{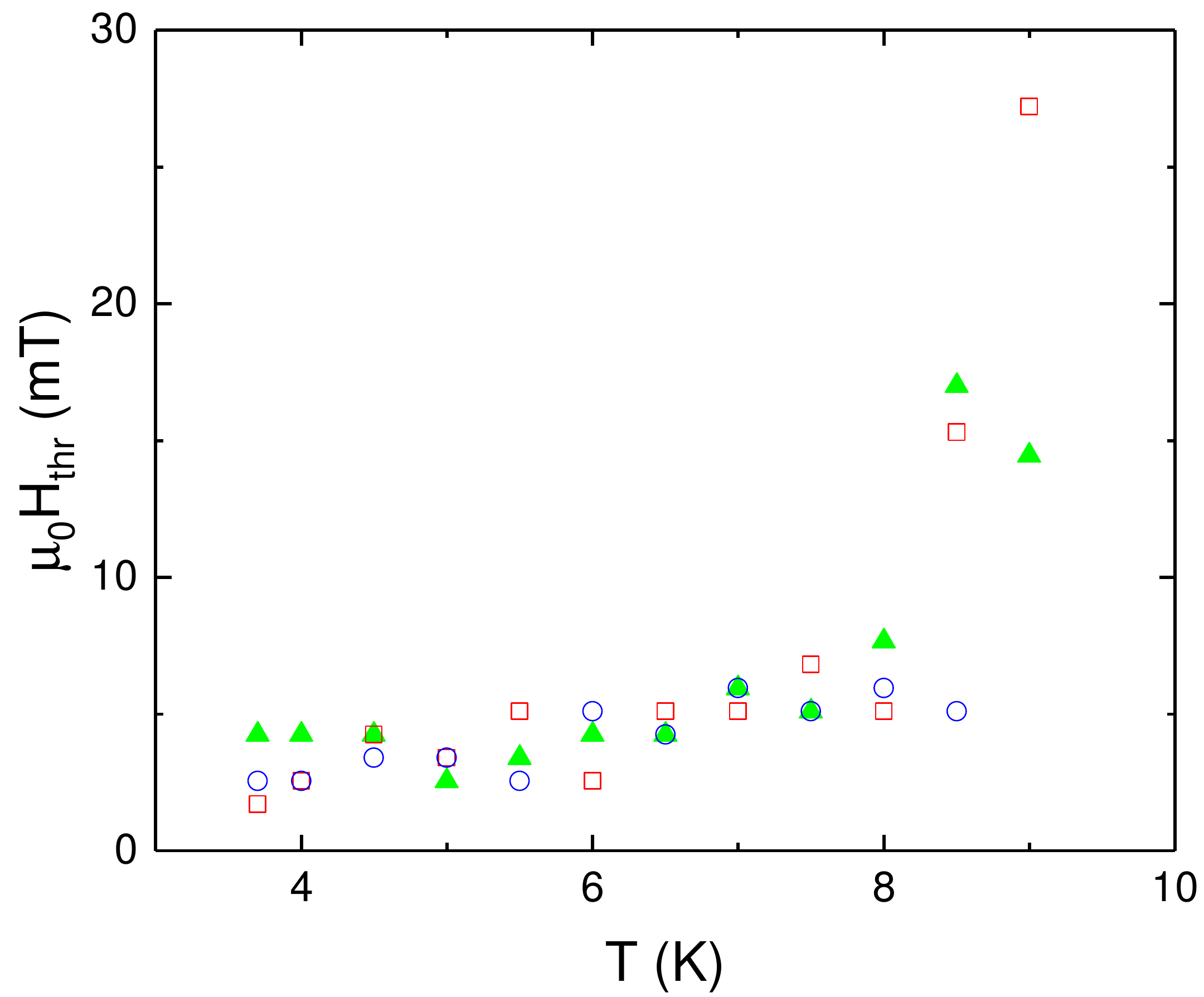}
	\caption{
		\label{fig:Bthr}
		The lower threshold field for the appearance of dendritic avalanches as a function of temperature. At a given temperature, the red square, green triangle and blue circle correspond to the field of appearance of the first dendritic avalanche in three separate experiments. 
	}
\end{figure}

The fields at which the first dendritic avalanche appeared at a given temperature were not the same in the three consecutive experiments. The lower threshold field $\mu_{\rm 0}H_{\rm thr}$ for the appearance of dendritic avalanches as a function of temperature $T$ is shown in figure~\ref{fig:Bthr}. The temperatures are 3.7 K and every 0.5 K from 4.0 to 10.0 K. At a given temperature, the red square, green triangle and blue circle correspond to the fields at which the first dendritic avalanche appeared in three different MOI experiments. Dendritic avalanches were observed in all three experiments at temperatures up to 8.5 K. At 9.0 K a dendritic avalanche was only observed in two of the experiments, and none were observed at 9.5 or 10.0 K. Since the applied magnetic field, limited by 85 mT, was far below the field needed for full penetration at the temperatures of the experiments, we cannot determine exact threshold temperature $T_{\rm thr}$ for the disappearance of dendritic avalanches, but can state that it is above 9.0 K. In figure~\ref{fig:Bthr}, $\mu_{\rm 0}H_{\rm thr}$ is merely constant at low temperatures, but increases a little with increasing $T$ up to 8.0 K, and increases rapidly above 8.0 K. This relationship between $\mu_{\rm 0}H_{\rm thr}$ and $T$ is similar to what has been previously reported for 300-nm thick films of MgB$_2$ \cite{PhysRevLett.97.077002} and 500-nm thick Nb films \cite{Welling200411}, but the rapid increase occurs at different temperatures in those experiments. The variance of $\mu_{\rm 0}H_{\rm thr}$ is also merely constant up to $T$ = 8.0 K and then increases, which can be seen for 8.5 and 9.0 K in comparison with that at lower temperatures. From theoretical models \cite{PhysRevLett.97.077002,PhysRevB.93.174511} one would expect the threshold field to be much higher in tapes. This indicates that the edge properties are determining the conditions for onset of the first avalanche. The threshold fields found from MOI experiments presented in figure~\ref{fig:Bthr} are much lower than 145 mT found from the $m$($H$) measurements at 5 K in figure~\ref{fig:mag}(a), because one could not see the smallest avalanches in magnetic moment measurements. If the magnitude of magnetic moment in $m$($H$) plots is lower than expected in areas where no flux jumps are seen, it is possible that there are in fact several flux jumps too small to be detected in such measurements, and being interpreted by the instrument as noise. MOI, although limited in operating range to low magnetic fields, can be employed to confirm the existence of such avalanches and visualize their propagation into the sample. 

\section{Conclusions}

Flux jumps caused by thermomagnetic instability have been investigated in an MgB$_2$ tape by MOI and magnetic moment measurements. Dendritic avalanches were observed with MOI and occurred at much lower magnetic fields than flux jumps seen in measurements of magnetic moment. The smallest thermomagnetic instabilities seen in MOI could not be observed in our magnetic moment measurements. The dendritic avalanches in the tape have similar properties to those appearing in thin films, but have relatively few branches and their patterns are in some cases partially reproducible. Most of them, however, are non-reproducible. The average size of new dendritic avalanches, as well as their branching, increases with increase of applied magnetic field. Upon increasing the temperature, the number of avalanches decreases. The lower threshold magnetic field for their appearance first increases slowly with increasing temperature and then increases rapidly above 8 K. The lower threshold field is much lower than what is expected from theoretical models, indicating that the onset is dictated by the edge properties of the tape.

\ack
This work was financially supported by the University of Oslo, the Spanish Ministerio de Econom\'{i}a y Competitividad, the European FEDER Program (Projects MAT2011-22719 and ENE-2014-52105-R) and the Gobierno de Arag\'{o}n (research group T12). The authors would like to acknowledge the use of Servicio General de Apoyo a la Investigaci\'{o}n-SAI, Universidad de Zaragoza and to thank I. Cabistany and J. A. G\'{o}mez for technical assistance with manufacturing the tapes. 

\section*{References}
\bibliography{bibTQ}

\end{document}